# Obtaining Traffic Information by Urban Air Quality Inspection

P.Ferante[1], D. Lo Bosco[2], S. Nicolosi[1], G. Scaccianoce[1], M. Traverso[1], G.Rizzo[1]

1*DREAM, Università degli studi di Palemo,*
2*DIMET, Università Mediterranea di Reggio Calabria,*
e-mail: *nicolosi@dream.unipa.it*

**ABSTRACT**

The level of air quality in urban centres is affected by emission of several pollutants, mainly coming from the vehicles flowing in their road networks. This is a well known phenomenon that influences the quality of life of people.

Despite the deep concern of researchers and technicians, we are far from a total understanding of this phenomenon. On the contrary, the availability of reliable forecasting models would constitute an important tool for administrators in order of assessing suitable actions concerning the transportation policies, public as well private.

As matter of fact, the definition of a physical model requires the knowledge of many parameters, involving the running fleet, the microclimatic conditions and the behaviour of driving people: the resulting complexity of such kind of informations (that also depend on standard and rules) makes almost impossible the assessing and the solution of these models.

In turn, in the recent years, a new approach is achieving, increasing popularity, which is essentially based on a phenomenological approach.

In other words, the physics of the phenomenon concerning the air pollution is neglected, while the attention is focused on the functional relationships between sources of pollution and local concentrations.

Such kind of approaches is made possible by means of the application of analytical models utilising the fuzzy logic, generally supported by genetic and/or neuronal algorithms.

Referring to the situation of the running fleet and the measured pollutant concentrations concerning the Italian town of Palermo, a *data-deduced* traffic model is here derived, its truthfulness being justified by a fuzzyfication of the phenomenon.

A first validation of the model is supplied by utilising the emissions characteristics and the pollutant concentrations referring to a two years period of time.

This work could represent a first attempt in defining a new approach to the problem of the pollution of the urban contexts, in order of providing administrators with a reliable and easier tool.

## 1. INTRODUCTION

Air pollution in urban centres comes mainly from the emissions of running fleet in road network. This well accepted statement has brought a lot of researchers to look for an analytical relationship linking the sources emissions with the local measured values of pollutant concentrations. Several modelling methodologies have evolved over the last 30 years (for a review see [1,2]),[3], but the complexity of the problem, mainly given by the large number of the variables involved, other than the vehicular fleet composition, makes the resulting models not easily solving; in other cases, the models descend from oversimplified hypotheses that often make the found solution not exportable in different urban contests: in this case the solution can be assumed as merely formal.

Moreover, researchers activity is also affected by the lack of data concerning the emission sources. The number of the vehicles actually circulating is, really, not simple to be measured: as a consequence, this number is very often derived by traffic models calibrated with several variables (e.g. demographic, socio-economic, and so on [2]). The details of these input variables do offer a good flexibility in the modelling process, but the level of uncertainty associated with them, however, shows that a reliable prediction for long time scale is almost impossible.

Our work aims to provide a contribution in the overcoming of these problems proposing a new traffic model, able to calculate the number of vehicles running in a period of time of one hour, once the ground level CO concentration is known. Further, on the basis of this traffic model and also keeping into account the *fuzzified* frame in which the measured concentrations are supposed to depend on temperature and solar radiation, the phenomenological relationship between local concentrations and emission sources is also justified.

This aim is achieved by implementing the neuro-fuzzy network supported by the Matlab Toolbox© called "ANFIS" and by characterizing it by means of three inputs (temperature, solar radiation and number of vehicles) and one output, that is the concentrations obtained by the measured values. The initial rules (relating input and output), the learning process (based on backpropagation and hybrid algorithm) and the proper validation of the model, have been implemented using the set of data of years 1997 and 2002. The relationship between CO concentration and the number of vehicles, shows that our traffic model is good enough, although it certainly requires further investigations.

## 2. TRAFFIC MODEL

Our approach in solving the dynamic behaviour of the puff of gas, coming out from a vehicle and diffusing in the urban environment, can be considered as totally phenomenological. The big number of data utilizable in order to follow the temporal behaviour of air pollutants, enables us to introduce some reasonable hypotheses about the system evolution.

The measured values of concentration depend on a multitude of variables, that anyway can be usefully divided in three groups. The first one contains the variables always affecting the local concentrations in the same way, discarding from the particular instant of time in which the measurement is performed: the urban configuration, for example. The second one is constituted by the set of local microclimatic conditions whose random variation modifies the concentration measurements. The third one is formed by the socio-demographic variables, whose influence can be related to the daily working-day traffic demand.

On the basis of this classification, the research of the phenomenological relationship between local concentration and all the previously cited variables can be done discarding the first group, whose influence will be entirely contained within some analytical constant parameter. The second group plays the major role, hardly influencing the local concentration. The third group is usually utilised to deduce quantitative schemes (traffic models) from the daily traffic demand. The main feature of these schemes is that the number and typology of vehicles running inside the urban network can be derived from socio-demographic data, as the annual census of population and housing, employment data and other relevant social input. The higher is the number of involved variables the higher is the reliability of the model. The interpretation frames in this way obtained can provide information about the average *per hour* number of the emission sources in an arbitrary working day.

Therefore, by accepting these models, we have implicitly assumed that every working day of the year is characterized by the same average *per hour* emissions. This hypothesis can be justified by a qualitative analysis of the hourly concentrations behaviour and it constitutes the basis for our data analysis. We will show that, on the ground of this assumption, a data-deduced traffic model can be built-up, so bypassing the problem of collecting socio-demographic data.

The average per hour measured values of pollutant concentrations in different receptor sites placed at ground level in the Italian town of Palermo show the same behaviour, as a function of the time, in all the working days of the year.

This not means that the measured values are the same but that they have maxima or minima in the same interval of time and the evolution between two adjacent opposite limit is almost of the same kind.

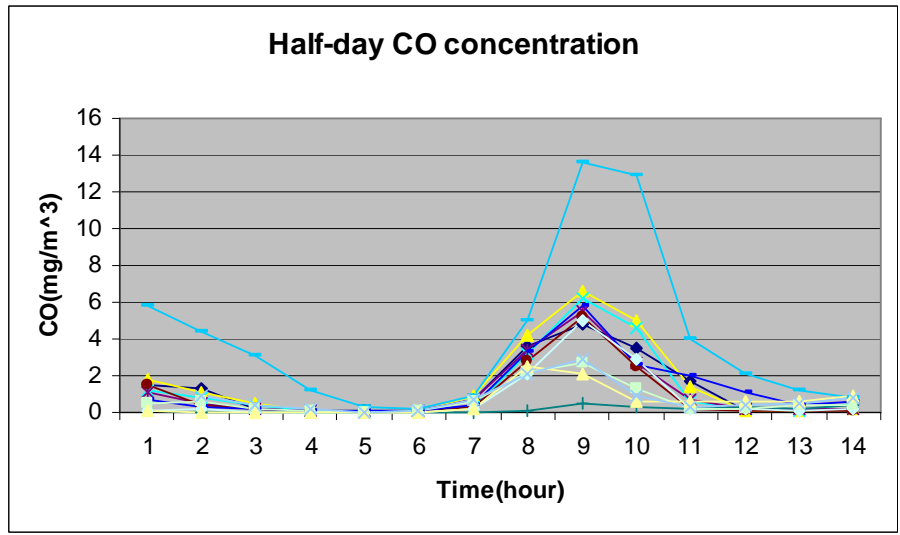

**Figure1. Tipical half day CO concentration**

Figure 1 displays the half-day average per hour concentration of CO measured in the central square of Palermo, that is "Piazza Castelnuovo", as a function of time in a few representative working day. The principal behaviour we can observe is that, from four o'clock to five o'clock in the morning, it takes its minimum value; then, it increases showing a maximum value at nine o'clock; subsequently, it decreases reaching a relative minimum at three o'clock p.m.

This feature induces to assume that the traffic demand doesn't depend on a particular working day, but only on the specific instant of time (of an arbitrary day) in which the measurement is taken. Moreover, it can be argued that the differences that can be found in the measured concentrations for the same site- in different days should be related to the local microclimatic conditions characterizing the chosen site. In other words, we can say that for fixed interval of time (one hour) there exists a constant parameter, namely the hourly average emission, that is attributable to the polluting sources insisting on to the "area of influence" constituted by the neighbourhood of the observation site.

Moreover, this constant emission supplies an upper limit for the ground level concentrations, since the other affecting-measurement variables can only cause a decrease in the pollutant concentration, whose maximum value is imposed by the entity and emissions of present sources [1].

Let us suppose now that all the random variables, except the wind speed, are constant and equal to the values for which the actual ground level concentration is the maximum one. If we could perform a measurement of the concentrations, under these ideal conditions, we would expect to find a set of points belonging to the curve that represents the functional dependence of the concentration from the wind speed. If, now, we think to leave the other parameters that are free to change, but for the sources emission (that is maximum from 8 to 9 a.m.), we would find in the same plane a cloud of points confusedly distributed inside a limited area, whose upper bound is given by the curve first identified.

---

[1] This isn't rigorously correct because it discards the correlation between present and past values, that can produce an accumulation of pollutants and, as a consequence, a measured value of concentration greater than those due to instantaneous performances of the sources. Anyway, we can assume that diffusive phenomena in some sense cause the concentration to loose "memory" of its past values.

This simple reasoning shows that we can obtain the analytical relationship linking concentrations and wind speeds under the ideal, previously identified, conditions, and that it has to be represented by the boundary line, providing the separation between the areas with and without points.

Further, if our arguments are corrects, changing the observation time or, equivalently, varying the number of emission sources, the cloud of points should be characterised by the same features with a boundary line that is lower than the previous one.

Figure 2 shows the overlapping of different series, corresponding to different intervals of time (from 3 to 9 a.m.), of the ground-level hourly average measured values of CO concentration [$C(\tau, z=0) \equiv C(\tau)$] as a function of the hourly average wind speed measured values. The figure appears to be in accordance with our arguments and, besides, the overlapping of the different coloured series singles out the family of *iso-emissive* curves. From these curves it appears that, increasing the wind speed, the concentration decreases following an exponential wise law. This behaviour is reasonable enough: in fact, when the value of the wind speed is zero the concentration takes its maximum allowed value; when the speed goes to infinity, concentration goes to zero discarding all other parameters.

The figure 3 shows the same concentration data in a logarithm scale.

The isoemissive curves are straight lines of the same slope and different *intercepta*, whose equation is the following:

$$Ln(C(\tau_i)) = C_{MAX}(\tau_i) - kV , \qquad i=1,...,24 \qquad (1)$$

or, equivalently,

$$C(\tau_i) = C_{MAX}(\tau_i) e^{-kV} \qquad (2)$$

with k≈3/2 as deducible by a linear fitting of figure 3.

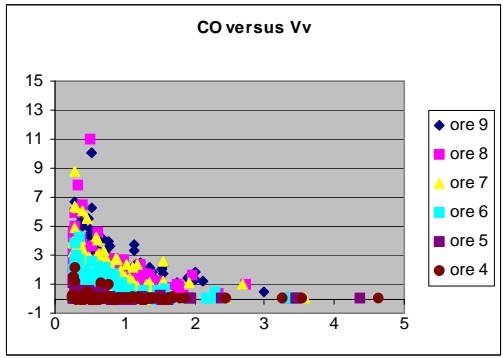
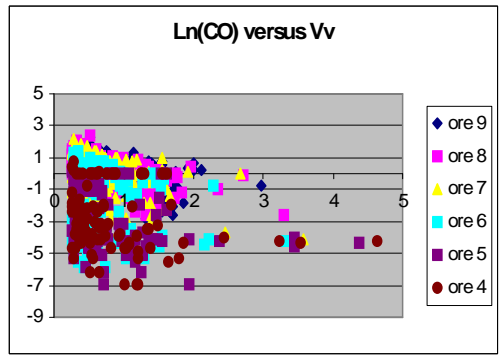

**Figura 2**            **Figura 3**

The times $\tau_i$ are discrete intervals of one hour amplitude, so that $\tau_i$ is the interval of time ranging from the hour *(i-1)-th* and the hour *i-th*, *i=1,...,24*. The constant *k* depends on the urban configuration, while $C_{MAX}(\tau_i)$, that is the hourly average value of concentration at the ground level corresponding to a windless condition, has to be related to the analyzed pollutant and to the number of vehicles running in the same interval of time. In order to find the analytical relationship between $C_{MAX}(\tau_i)$ and the number of emission sources, we can assume, for the sake of simplicity, that the sources show an uniform distribution in the (xy) plane (ground level).

This hypothesis, along with the definition of the "yearly average vehicle" (YAV) [4,5,6] of a whole modality of transport, does ensure that the pollutants' concentration doesn't depend on the x and y coordinates, so that the main diffusion (turbulent motion) of pollutants is only in the z direction.

Moreover, discarding the friction with the building wall, this turbulent motion causes the volume of pollutants to increase without suffering of any deformation. This means that the strain tensor *e* has only one non-zero component that is $e_{zz} = \frac{\partial v_z}{\partial z}$, where $v_z(z,t)$ is the pollutant velocity. The diffusion along z axis is a free expansion of an ideal gas under the influence of the gravitational field and of the temperature gradient. In this case the Navier Stokes equation assume the form [3,7]:

$$\frac{RT}{m}\frac{\partial \rho}{\partial z} + \frac{\rho R}{m}\frac{\partial T}{\partial z} + \rho g = -\rho\left(\frac{\partial v_z}{\partial t} + v_z\frac{\partial v_z}{\partial z}\right) - v_z\left(\frac{\partial \rho}{\partial t} + v_z\frac{\partial \rho}{\partial z}\right) + \frac{4\partial}{3\partial z}\left(\mu\frac{\partial v_z}{\partial z}\right) + Sv_z, \quad (3)$$

where $\rho(z,t)$ is the concentration of the analyzed pollutant, $m$ is its molecular mass, $R$ is the constant of ideal gases, $T$ is the absolute temperature, $g$ is the gravitational acceleration of Earth. Finally:

$$\left(\frac{\partial \rho}{\partial t} + v_z\frac{\partial \rho}{\partial z}\right) = S \quad (4)$$

is the equation of continuity linking the variation of volumetric pollutant mass to the volumetric emission rate $S$ from pollution sources.

The calculation of the emission can be done referring to the pollutant emission factors provided by the COPERT III European method [8], along with the definition of the "yearly average vehicle" emission factor (*YAV*) [4,5,6] of a whole given modality of transport. In this way the volumetric emission rate can be obtained by the following equation

$$\iint S\partial z\partial p_\perp = \frac{\partial n(t)}{\partial t} YAV, \quad (5)$$

where $n(t)$ is the number of running vehicles at the time instant $t$; now, indicating with $\partial p$ the elementary route follow by the YAVs, $\partial p_\perp$ has to be interpreted as its orthogonal direction. Therefore, the total emitted pollutant mass in the time interval $\tau_i$ is:

$$M_E(\tau_i) = \int_{t_0}^{t}\partial t\int \partial z\int \partial p\int \partial p_\perp S = N(\tau_i)\int YAV\partial p = N(\tau_i)YAV<p>, \quad (6)$$

where $N(\tau_i)$ is the number of vehicles that have run in the one hour amplitude interval $\tau_i$ and $<p>$ is its average route.

As already highlighted, from 4 o'clock to 5 o'clock in the morning the registered level of concentration are so much low that it is reasonable to assume that, in this interval of time the concentration may reach an equilibrium configuration, because of the low number of sources and the fastness of turbulent diffusion processes (whose characteristic time can be assumed much lower than one hour). Under this hypothesis and assuming that the temperature is steady state during the analyzed hour, the Navier Stokes equation can be easily solved by putting the right member of eq.(3) equal to zero, so obtaining:

$$\rho(z,5) = \rho(0,5)\frac{T(0,5)}{T(z,5)} e^{-\int_0^z \frac{mg}{RT(z',5)}\partial z'}. \quad (7)$$

The above equation can be integrated over the total free space in order to calculate the global pollutant mass emitted by the $N(\tau_5)$ running vehicles, so obtaining the total mass equation, in the form:

$$M_E = A\rho(0,5)T(0,5)\int_0^\infty \frac{e^{-\int_0^z \frac{mg}{RT(z',5)}\partial z'}}{T(z,5)}\partial z, \quad (8)$$

where $A$ is the area of the influence assumed to have the same size of the surface of Piazza Castelnuovo.

Let us assume that:

$$\begin{cases} T(z,t) = T(0,t) + a(t)z & 0 \leq z \leq h_1 \\ T(z,t) = T(0,t) + b(t)z & h_2 \leq z \leq h_3 \\ T(z,t) = T(0,t) + c(t)z & h_3 \leq z \leq h_4 \\ \vdots & \vdots \end{cases} \quad (9)$$

with $a(t), b(t), c(t),... \in \Re$. The previous one is a general enough atmospheric configuration able to reproduce the variation of temperature in the atmospheric boundary layer with height during day and night because of the dependence on time of the temperature gradients. Let us consider, for example, the atmospheric structure proposed by Jacobson, M. Z. (1999)[9].

The boundary layer depth ranges between 500 and 3000 m and it is divided into the surface layer (the first 10% closest to the ground) and the neutral convective layer, also called the mixed layer. At the top of the boundary layer, we have the inversion layer. This structure is shown on the figure 4.

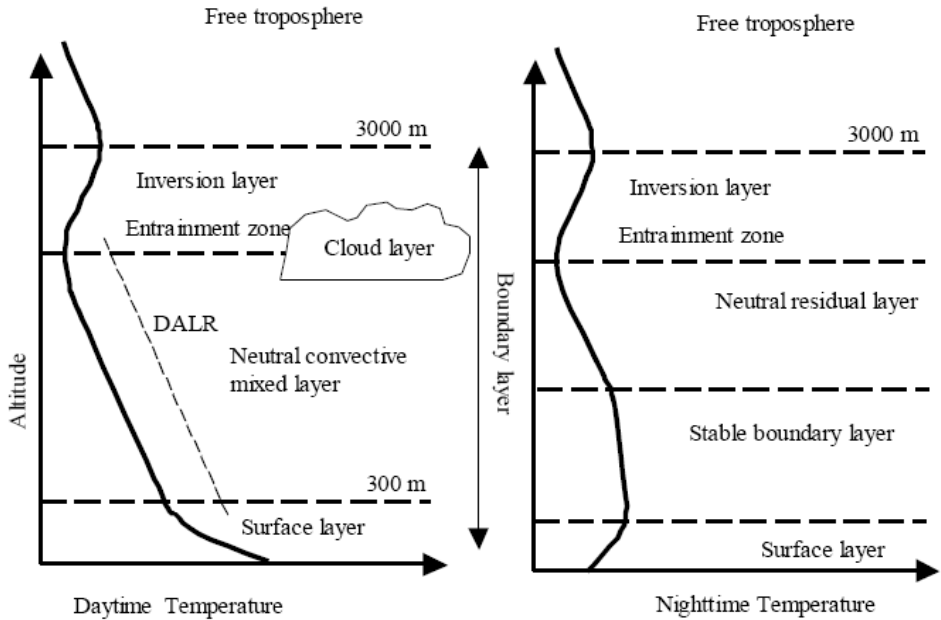

**Figure 4. Variation of temperature in the atmospheric boundary layer with height during day and night (from Jacobson, 1999).**

The vertical temperature gradient is different during the day and at night. At night, the surface layer and a large part of the mixed layer become stable as the temperature increases with height due to radiative cooling of the ground. At the top of the boundary layer, the temperature gradient is usually positive (i.e. stable) and the turbulence dies there. This is the inversion layer. Between the inversion and the neutral layers, we have the so-called entrainment zone, which is where we see often cloud formation. Because it is not easy to penetrate the inversion layer, it can be considered as the ceiling of pollution. Therefore, most of the pollution emitted at ground level will disperse approximately up to the first 3 km of the atmosphere, with further diffusion to the free troposphere being a slower process.

It is not difficult to convince oneself that eq. (9) well represents the atmospheric structure previously described. Moreover, it permits the integration of the total mass equation leading to

$$M_E(5) = \frac{A\rho(0,5)T(0,5)R}{mg}. \quad (10)$$

Inserting eq.(6) in the previous one, we obtain a linear equation for the unknown quantity $N(\tau_5)$ that can be solved. We obtain:

$$N(\tau_5) = \frac{A\rho(0,5)T(0,5)R}{mg(YAV)<p>} \quad (11)$$

The knowledge of this number is useful to obtain the number of vehicle $N(\tau_i)$ that run from *(i-1)-th* to the *i-th* hour or, equivalently, the working day traffic model.

Actually, we are not able to solve the Navier Stokes equation in an arbitrary interval of time, because the dynamical processes aren't negligible. Notwithstanding, it appears reasonable to assume that $C_{MAX}(\tau_i)$ is proportional to the global emission appeared in the *i-th* hour, reduced by an unknown scale factor $f(X(i),Y(i),…)$ depending on the local microclimatic variable $X(i),Y(i),…$ , with *i=1,…,24*, that is:

$$C_{MAX}(\tau_i) = N(\tau_i)(YAV)f(X(i),Y(i),...), \quad 0 \leq f(X(i),Y(i),...) \leq 1 \quad (12)$$

If, in addition, we assume that $X(i) \approx X(i-1)$, $Y(i) \approx Y(i-1)$,…then $f(X(i),Y(i),…) \equiv f(X(i-1),Y(i-1),…)$. On the ground of this hypothesis it is immediate to convince oneself that:

$$\frac{C_{MAX}(\tau_i)}{C_{MAX}(\tau_{i-1})} = \frac{N(\tau_i)}{N(\tau_{i-1})} \quad (13)$$

and that:

$$N(\tau_i) = \frac{C_{MAX}(\tau_i)}{C_{MAX}(\tau_5)} N(\tau_5). \quad (14)$$

A statistical analysis of the calculated $N(\tau_i)$ shows that the frequency of appearance of each $N(\tau_i)$ follows a Gaussian distribution. It is, indeed, reasonable to take the mean values of the calculated sets of vehicular number as the best estimation of the average per hour number of vehicles and its standard deviation as the associated error. In this way we obtain:

$$<N(\tau_9)> = 16590 \pm 7760$$
$$<N(\tau_8)> = 13940 \pm 6170$$
$$<N(\tau_7)> = 6860 \pm 3150$$
$$<N(\tau_6)> = 2270 \pm 1420$$
$$<N(\tau_5)> = 1190 \pm 1050$$

The traffic model we have generated is entirely based on hypotheses that cannot be validated in a direct way. For this reason, in the next session we will try to develop arguments that, in some respect, give a justification of the truthfulness of our traffic model.

**3. FUZZY LOGIC**

The analysis developed inside the traffic model section, starting from theoretical assumptions, discards the complex dependence of the measured value of concentration from the other local microclimatic variables, only taking into account its relationships to the wind speed. In this way we have introduced a traffic model able to give information about the number of circulating vehicles, once the hypotheses to make it working are satisfied. Unfortunately we haven't the possibility to test our traffic model by a direct inspection of the traffic flow.

Despite this fact, our aim is to supply an useful tool to local administration that is able to provide the actual traffic situation, so that pathological conditions in the measured value of concentration could be related to traffic condition supplying the basis for a decision-making process.

In general, the measured concentration will take a value belonging to the statistical mixture of all permitted values, although conditioned by the actual value of the random microclimatic conditions.

Unfortunately, apart from the wind speed dependence on the local concentration, it is not simple to explain how the other microclimatic variables analytically influence its value. This is due, primarily, to the incompleteness and imprecision of our data, at this stage.

In order to represent and manipulate incomplete databases, several extensions of the classical relational model have been proposed. Particularly, the fuzzy set theory [10], referring to the fuzzy logic proposed by

Zadeh [11,12], provides a suitable mathematical framework for dealing with incomplete and imprecise information.
Therefore the possibility of singling out the analytical relationship linking the average of the *per hour* windless pollutant concentration to the local microclimatic conditions, as well as to the local vehicular fleet, is here supplied by means of a fuzzyfication of the equation (12):

$$C_{MAX}(\tau_i) = N(\tau_i)(YAV) f(X(i), Y(i),...) \quad (15)$$

This equation is one of the ground hypotheses on which the traffic model is built on.
The choice of the inputs $X(i), Y(i)...$ is based on a proper account of "*where the pollutant went*" and "*what happened to it*". Assuming that
- the pollutant is released to the environment by the *turbulent dispersion* flow due to the atmospheric structure (vertical temperature distribution);
- the pollutant undergoes photochemical transformations;
- the pollutant level is increased by emissions from pollutant sources;

it appears reasonable to choose the ground level temperature, the solar radiation and the previously calculated number of vehicles as inputs for the fuzzy network.

## 4. RESULTS

The first step of the fuzzyfication process consists in the assignment of the training and checking data sets. The lack of an objective criterion in choosing these data sets and starting from the availability of the fuzzy network to represent the pollutant behaviour in the years 1997 and 2002, has induced us to make a mixing of these sets in order of avoiding a preferential role. The distribution of the training and checking data sets, as measured, is shown in the figure 5.

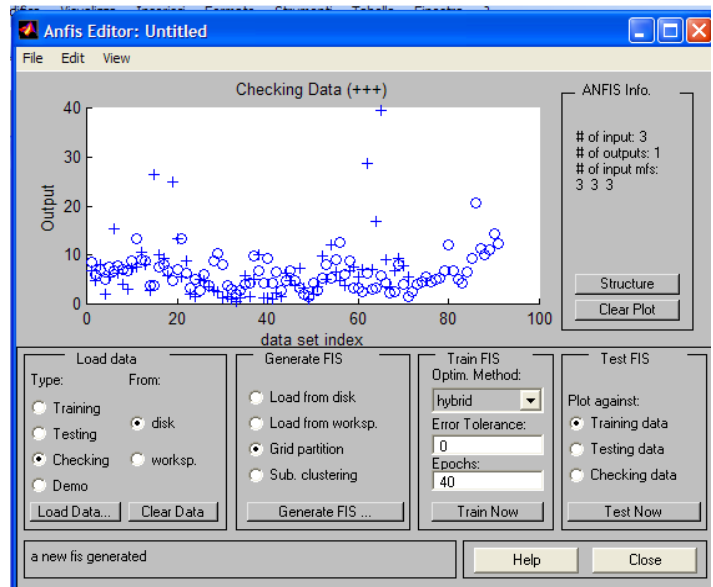

**Figure 5. The distribution of training and checking data sets.**

The assignment of the membership functions to the input variables and the definition of rules linking inputs and output has been made on the ground of the observation of the graphs depicting the pollutant as a function of temperature and solar radiation.
The comparison between the theoretical output and the measured ones has introduced an error whose minimization is provided by means of a hybrid algorithm (a combination of backpropagation and least-squares method) resulting in a correction of the previous assignments (membership functions and rules).
After about 120 epochs of training the neural fuzzy network has reached a steady value of the error (*e=0,07*) and has produced the distribution of training test data displayed in figure 6.

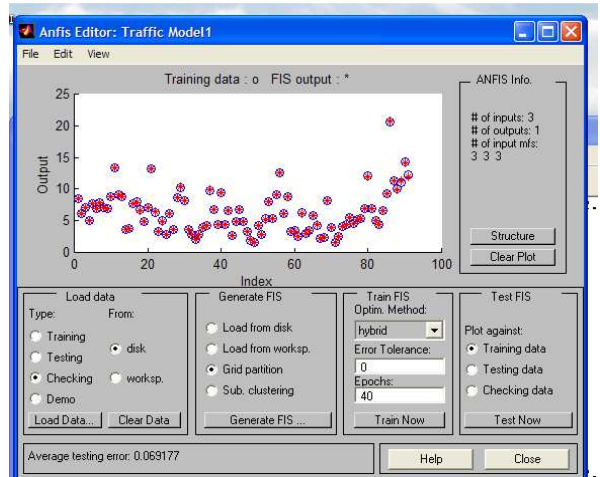
**Figure 6. Distribution of the trading test**

The small error, along with the full overlapping of the measured data with the theoretical output, ensures that the learning process has been correctly implemented. Moreover, the neural adaptive fuzzy interference system, whose principal structure is shown in the figure 7, has produced the analytical dependence of windless concentration as a function of the vehicular fleet showed in figure 8

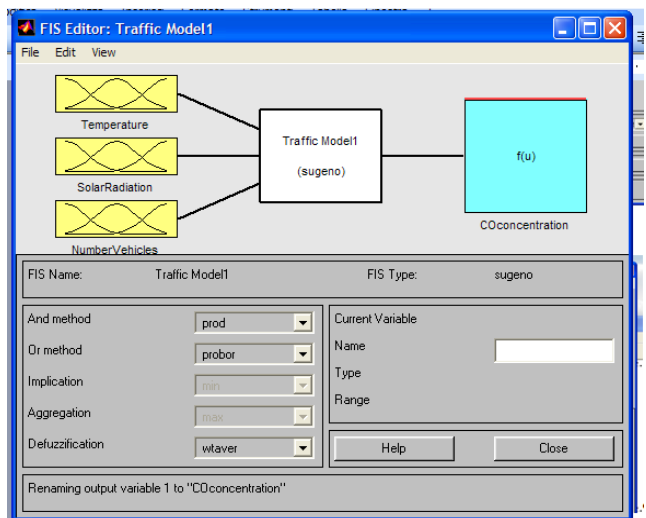
**Figure 7. Fuzzy inference system of traffic model**

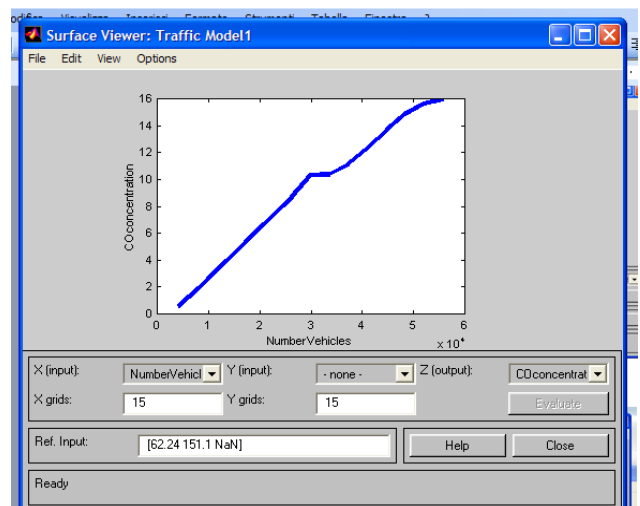
**Figure 8. The relationship between number of vehicles and CO concentrations**

The last figure predicts that the CO concentration increases as the number of running vehicles rises. A deviation from this behaviour occurs if this number is grater than 30.000 vehicles. In this regime the chaotic and dynamical diffusive phenomenon probably invalidates the ground hypotheses upon which the model has been derived.

**5-CONCLUSIONS**

This work takes its place in the wide research activity concerning the air quality level in urban centres.
Its main result is the implementation of a traffic model able to relate the measured values of CO concentration to the running fleet of vehicles that, as it is well known, can be considered the main source of urban pollution.
The method here developed, although requiring further investigations, contains the novelty of deducing the average *per hour* number of vehicles on the basis of a few reasonable hypotheses.
The truthfulness of this hypotheses probably needs major speculative arguments, obviously. Anyway, the reasonability of the first outcome of the present *data-deduced* traffic model, clearly encourage to better investigate where the hypotheses could fall down, in the aim of producing a predictive tool to support local administrators in the decision-making processes.